\documentclass[reprint,prx,twocolumn,twoside,aps,groupaddress]{revtex4-2}

\usepackage{amsmath,amssymb,amsfonts}
\usepackage{graphicx}
\usepackage{siunitx}
\usepackage{natbib}
\usepackage{braket}
\usepackage{multirow}
\usepackage[normalem]{ulem}
\usepackage{xspace}
\usepackage{xcolor}
\usepackage{url}
\usepackage[breaklinks]{hyperref}
\usepackage{braket}
\usepackage{lipsum}

\hypersetup{
    unicode=true,
    colorlinks=true,
    linkcolor=blue,
    citecolor=blue,
    urlcolor=blue
  }
\usepackage{breakurl}

\newcommand{\feshbach}{$\ket{F}$\xspace}
\newcommand{\s}{$\ket{s}$\xspace}
\newcommand{\ground}{$\ket{G}$\xspace}

\newcommand{\intermediate}{$\ket{E}$\xspace}

\newcommand{\atompairgrounddetail}{${(f=1,m_f=1)}_\mathrm{Rb}{(f=3,m_f=3)\xspace}_\mathrm{Cs}$\xspace}

\newcommand{\sigmapm}{$\sigma^{\pm}$\xspace}
\newcommand{\sigmap}{$\sigma^{+}$\xspace}
\newcommand{\sigmam}{$\sigma^{-}$\xspace}

\usepackage{xr}
\begin{document}
\title{Enhanced quantum control of individual ultracold molecules using optical tweezer arrays}

\newcommand{\physics}{Department of Physics, Durham University, South Road, Durham, DH1 3LE, United Kingdom}
\newcommand{\jqc}{Joint Quantum Centre Durham-Newcastle, Durham University, South Road, Durham, DH1 3LE, United Kingdom}

\author{Daniel K. Ruttley}
\affiliation{\physics}
\affiliation{\jqc}
\author{Alexander Guttridge}
\affiliation{\physics}
\affiliation{\jqc}
\author{Tom R. Hepworth}
\affiliation{\physics}
\affiliation{\jqc}
\author{Simon L. Cornish}
\email{s.l.cornish@durham.ac.uk}
\affiliation{\physics}
\affiliation{\jqc}

\begin{abstract}
Control over the quantum states of individual molecules is crucial in the quest to harness their rich internal structure and dipolar interactions for applications in quantum science.
In this paper, we develop a toolbox of techniques for the control and readout of individually trapped polar molecules in an array of optical tweezers.
Starting with arrays of up to eight Rb and eight Cs atoms, we assemble arrays of RbCs molecules in their rovibrational and hyperfine ground state with an overall efficiency of 48(2)\%.
We demonstrate global microwave control of multiple rotational states of the molecules and use an auxiliary tweezer array to implement site-resolved addressing and state control.
We show how the rotational state of the molecule can be mapped onto the position of Rb atoms and use this capability to readout multiple rotational states in a single experimental run.
Further, using a scheme for the mid-sequence detection of molecule formation errors, we perform rearrangement of assembled molecules to prepare small defect-free arrays. Finally, we discuss a feasible route to scaling to larger arrays of molecules.

\end{abstract}

\date{\today}

\maketitle

\section{Introduction}

Ultracold molecules offer a versatile platform for quantum science \cite{Carr2009,Bohn2017,Softley2023}, with applications spanning quantum simulation \cite{Barnett2006,Gorshkov2011,Baranov2012,Lechner2013,Wall2015,Yao2018} and quantum information processing \cite{DeMille2002,Yelin2006,Zhu2013,Ni2018,Hughes2020,Albert2020,Sawant2020} to ultracold chemistry \cite{Krems2008,Heazlewood2021,Liu2022} and precision measurement \cite{Schiller2005,Chin2009,Hudson2011,Andreev2018,Roussy2023}. Molecules feature a rich internal structure, constituting a ladder of rotational states with long radiative lifetimes. 
Preparing molecules in a superposition of rotational states engineers dipole-dipole interactions that can be controlled using microwave fields. These properties make rotational states well-suited for applications as qubits \cite{Bao2023a,Holland2023} or pseudo-spins in a quantum simulator \cite{Yan2013,Li2023,christakis2023}.
Moreover, the abundance of long-lived rotational states unlocks possibilities such as synthetic dimensions in the rotational degree of freedom \cite{Sundar2018}, realization of qudits \cite{Sawant2020} or the implementation of quantum error-correcting codes in the molecule's internal states \cite{Albert2020}.

Realization of many of these theoretical proposals demands a high level of control of the quantum states of individual molecules. Attaining such control is pivotal to exploit the wide array of tools offered by ultracold molecules for quantum science. Significant progress has been made in preparing and manipulating internal and external states of molecules \cite{Langen2023,Cornish2024} but control and detection of individual molecules in a single internal and external quantum state is an ongoing challenge.

Optical tweezer arrays are a powerful platform for the trapping, control, and readout of single ultracold particles \cite{Schlosser2001,Schlosser2002,Kaufman2021}. Arrays of tweezers are dynamically reconfigurable, allowing flexible connectivity \cite{Bluvstein2022} and enabling the preparation of states with low configurational entropy through rearrangement of particles \cite{Lee2016c,Barredo2016,Endres2016}. In this tweezer array platform, long-range interactions between trapped particles have been utilized to simulate complex quantum systems \cite{Labuhn2016,Bernien2017,Browaeys2020}. The platform's inherent scalability \cite{Evered2023} provides a promising avenue for constructing arrays with an even greater number of particles.

\begin{figure*}
\includegraphics[width=\hsize]{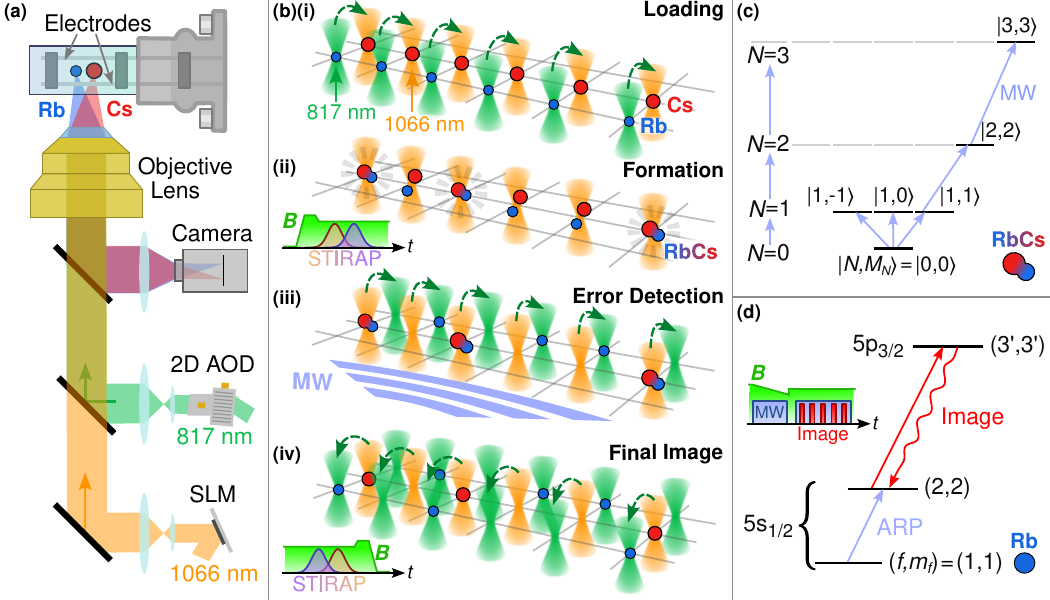}
\caption{Overview of the experimental apparatus and methodology for creating, controlling and detecting ultracold RbCs molecules. (a) The simplified setup showing the vacuum cell, objective lens and key elements of the optical setup. Arrays of 1066\,nm tweezers are created with a spatial light modulator (SLM). Arrays of 817\,nm tweezers are created with a two-dimensional acousto-optic deflector (2D AOD). Trapped atoms are detected by imaging atomic fluorescence onto a camera. (b) Stages of a typical experiment. (b)(i) Initially Rb and Cs atoms are loaded and rearranged to prepare defect-free 1D arrays in species-specific tweezers. (b)(ii) The Rb tweezers are merged to overlap with the Cs tweezers. The atom pairs are magnetoassociated and the resultant molecules are transferred to the ground state using Stimulated Raman Adiabatic Passage (STIRAP), as illustrated in the inset. (b)(iii) Atom pairs remaining due to failed molecule formation are separated. The Cs is ejected and the Rb is stored in a separate row of tweezers. Detection of the Rb atom indicates failure to form a molecule in a particular site. An experiment is performed on the molecules using microwaves to address rotational transitions. (iv) Finally the molecules are dissociated and the resulting atom pairs are separated into their original traps for imaging. (c) The lowest-energy rotational levels of RbCs labelled with the rotational angular momentum $N$ and its projection $M_N$. Arrows indicate the microwave transitions used in this work. (d) Energy levels used in imaging of Rb at high magnetic field. The closed cycle $(f=2,m_f = 2)\rightarrow(3',3')$ is used for imaging the atom at high field; the transfer $(1,1)\rightarrow(2,2)$ is performed with adiabatic rapid passage (ARP).}
\label{fig:overview}
\end{figure*}

The extension of tweezer arrays to ultracold molecules has been realized recently for both laser-cooled \cite{Anderegg2019,Holland2023b,Vilas2023} and assembled molecules \cite{Cairncross2021,Ruttley2023}. However, the full toolbox of techniques developed for ultracold atoms in optical tweezers has yet to be extended to the more complex molecular systems. In this paper, we address this gap by extending established experimental techniques demonstrated in neutral-atom tweezer arrays, which include rearrangement \cite{Endres2016,Barredo2016}, erasure conversion \cite{Scholl2023} and mid-circuit operations \cite{Graham2023,Huie2023,Lis2023,Norcia2023}, to apply them to ultracold bialkali molecules. Specifically, we globally and locally control multiple rotational states of individually-trapped ultracold molecules. 
We introduce a technique for the readout of multiple rotational states in a single iteration of the experiment, achieved by mapping onto atomic states, and demonstrate rearrangement of molecules using mid-sequence detection of formation errors.

The structure of the paper is as follows. Section~\ref{sec:overview} gives an overview of our experimental platform. Section~\ref{sec:formation} describes our procedure for the assembly of molecules in optical tweezers and reports the efficiency of this process. Section~\ref{sec:global} demonstrates global control of the rotational states of molecules in our optical tweezer array using microwave fields to perform coherent multi-photon excitation. Section~\ref{sec:readout} describes the detection of multiple rotational states of molecules in a single experimental run. Section~\ref{sec:local} demonstrates local control of rotational states using an addressing tweezer in combination with microwave fields to selectively excite specific molecules in the array. Section~\ref{sec:rearrangement} describes the detection of molecule formation errors and rearrangement of molecules to prepare a defect-free array. Finally, Section~\ref{sec:outlook} examines the prospects for scaling the techniques described in this paper to larger arrays.

\section{Overview of the experimental platform}\label{sec:overview}
Figure~\ref{fig:overview}(a) shows an overview of the experimental apparatus~\cite{Brooks21,Spence22} we use to produce ultracold $^{87}$Rb$^{133}$Cs (hereafter RbCs) molecules trapped in one-dimensional arrays of optical tweezers.
A key aspect of our experimental setup is the use of two distinct wavelengths of optical tweezers which enables species-specific trapping and independent control of the atoms and molecules.
Tweezers at a wavelength of 1066\,nm are attractive to all species in our experiment, whereas tweezers at 817\,nm are strongly attractive for Rb, weakly attractive for RbCs, and repulsive for Cs.
The 1066\,nm tweezers are created with a spatial light modulator (SLM) and the 817\,nm tweezers are created with a two-dimensional acousto-optic deflector (2D AOD). Both wavelengths are aligned through a high numerical aperture objective lens to generate the tweezers in an ultra-high vacuum glass cell. The SLM generates a static array whereas, by changing the radio-frequency tones applied to the 2D AOD, we can dynamically switch and move the 817\,nm tweezers mid-routine to manipulate the atoms and molecules.

Molecules are assembled from optical tweezer arrays of individually trapped Rb and Cs atoms [Fig.~\ref{fig:overview}(b)(i)] to form an array of RbCs molecules in the rovibrational ground state [Fig.~\ref{fig:overview}(b)(ii)].
As the molecules are individually trapped we avoid loss caused by molecular collisions \cite{Takekoshi2014,Gregory2019,Gregory2020,Gregory2021a} and are able to selectively control individual molecules.
The molecules formed occupy a single internal quantum state and the assembly from laser-cooled atoms produces molecules predominantly in the motional ground state of the optical tweezers. Microwave fields can then be used to manipulate the rotational state of the molecules [Fig.~\ref{fig:overview}(b)(iii)]. The transitions and states that we explore in this work are highlighted in Fig.~\ref{fig:overview}(c).

We exploit individual control of the species in our experiment to readout information.
Figure~\ref{fig:overview}(b)(iii) highlights a general scheme for the indirect detection of molecules, whereby an atom-specific tweezer array is used to pullout Rb atoms from an array of tweezers that is partially filled with molecules.
By measuring the occupancy of this ``detection array'', we can infer the lack of molecules in the corresponding traps in the primary array.
This detection scheme is discussed further in Sec.~\ref{sec:mol_det_and_form_eff}.
By imaging these atoms mid-routine using the scheme shown in Fig.~\ref{fig:overview}(d), we can detect molecule formation errors and, hence, perform rearrangement of the occupied traps, as discussed in Sec.~\ref{sec:rearrangement}."
Alternatively, by repeatedly converting different molecular states to atom pairs and using the pullout method discussed above, we can map the molecular states onto multiple detection arrays for multi-state readout, as described in Sec.~\ref{sec:readout}.

\section{Molecule formation}\label{sec:formation}
\subsection{Formation of weakly bound molecules}
All experiments begin by stochastically loading individual Rb and Cs atoms in arrays of optical tweezers at wavelengths 817\,nm and 1066\,nm, respectively.
The atoms are imaged with fidelity $>99.9\%$ \cite{Brooks21} and dynamically rearranged to create defect-free one-dimensional arrays with up to eight atoms of each species, as illustrated in Fig.~\ref{fig:overview}(b)(i).
Following this, the atoms are prepared predominantly in the motional ground state using Raman sideband cooling and transferred to the hyperfine state \atompairgrounddetail \cite{Spence22}.
The arrays are then merged along the inter-array axis to generate atom pairs confined in the 1066\,nm array. The merging is carefully optimized to minimize heating such that the atom pairs predominantly occupy the ground state of relative motion.
We find that the atom pairs are successfully prepared in the ground state of relative motion in 56(5)\% of experimental runs. The infidelity in the initial preparation of each species in the correct hyperfine state is approximately 1\% prior to merging.
However, significant Raman scattering of Rb caused by the 817\,nm tweezer during the merging process means that atom pairs in the correct hyperfine state are prepared in 93(2)\% of runs.
The remaining 7(2)\% of atom pairs are prepared in excited hyperfine states and, for the duration of typical experimental routines, are lost due to inelastic collisions following merging.

\begin{figure}
\includegraphics[width=\hsize]{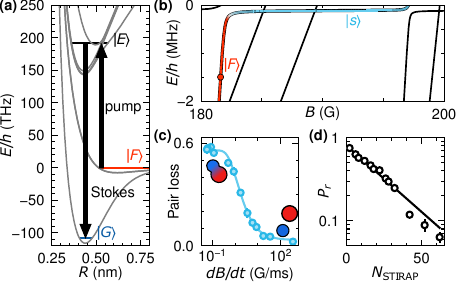}
\caption{(a) Electronic energy structure of RbCs molecules highlighting the STIRAP pathway connecting the Feshbach state \feshbach to the ground state \ground. (b) Near-threshold molecular bound states and the pathway for magnetoassociation. Atom pairs are associated into the near-threshold state \s (blue) by sweeping the magnetic field down across the Feshbach resonance at 197.1\,G. The magnetic field is then decreased further to transfer the molecules into state \feshbach (red) at $181.7$\,G. (c) Probability of losing an atom pair as a function of the magnetic field ramp speed $dB/dt$ over the avoided crossing at 197.1\,G. Molecule loss is induced by subsequently sweeping the magnetic field to 181.7\,G and applying pump light. 
(d) Efficiency of STIRAP as a function of the number of one-way STIRAP transfers, $N_\mathrm{STIRAP}$. We extract a one-way efficiency of 96.4(1)\% from the fitted solid line.}
\label{fig:molecule_formation}
\end{figure}

Molecule formation is achieved in two steps. We first utilize an interspecies Feshbach resonance to magnetoassociate atom pairs into molecules \cite{Zhang2020,Ruttley2023}.
The molecules are then transferred from the weakly bound state \feshbach to the rovibrational ground state \ground using two-photon stimulated Raman adiabatic passage (STIRAP) following the scheme illustrated in Fig.~\ref{fig:molecule_formation}(a).
The energy pathway used to access state \feshbach from the atom pair state is shown in Fig.~\ref{fig:molecule_formation}(b).
Starting at a magnetic field of 205.8\,G, atom pairs are adiabatically transferred into a near-threshold molecular state \s by ramping the magnetic field over an interspecies Feshbach resonance at 197.1\,G. The magnetic field is then lowered further to traverse an avoided crossing associated with a Feshbach resonance at $\sim182$\,G, transferring the molecules to state \feshbach at a field of 181.699(1)\,G.

Only atom pairs in the required hyperfine state and the ground state of relative motion can be magnetoassociated to form a molecule \cite{Busch1998,Zhang2020}. We measure the conversion efficiency from atom pairs to molecules in state \feshbach by inducing state-sensitive loss of molecules~\cite{Ruttley2023}.
By applying a 1\,ms pulse of resonant ``pump'' light at 1557\,nm, we excite molecules in state \feshbach to state \intermediate 
from which they may decay to other states by spontaneous emission.
When this happens we do not recover atom pairs following the reversal of the association ramps~\cite{Ruttley2023}. This allows us to optimize the parameters of the magnetoassociation sequence by measuring the probability to lose the atom pair.

Figure~\ref{fig:molecule_formation}(c) shows the result of such a measurement where we vary the magnetic field ramp speed $dB/dt$ during the magnetoassociation sweep across the Feshbach resonance at 197.1\,G.
At high $dB/dt$, the avoided crossing between the atom-pair and molecule states is traversed diabatically and molecules are not formed.
The pump light then has no effect and atom pairs are recovered at the end of the sequence in 97(1)\% of experimental runs. Here the background loss of 3(1)\% is from atom pairs that occupy an excited hyperfine state. This is lower than the 7(2)\% infidelity in the hyperfine state preparation mentioned above, as in this measurement the atom pairs are held for a shorter time such that the collisional loss does not saturate.
When a slower magnetic field ramp is used, we adiabatically transfer to \s then subsequently to \feshbach and molecules are lost once the pump light is applied.

The solid line in Fig.~\ref{fig:molecule_formation}(c) shows the result of fitting a Landau-Zener model to the data, where the probability of traversing the avoided crossing adiabatically is $p = 1-\exp({-(4\pi^2n_2\hbar/2\mu)|a_s\Delta/(dB/dt)|})$ \cite{Hodby2005,Koehler2006,Chin2010}.
Here $\mu$ is the reduced mass of the two atoms, $a_s = 645(60)a_0$ is the background s-wave scattering length and $\Delta = 90(10)$\,mG is the width of the Feshbach resonance \cite{Takekoshi2012}. 
From this model we extract the atom-pair density $n_2 = 1.7(1)\times 10^{13}$\,cm$^{-3}$ and fit an atom pair to molecule conversion efficiency of 53(1)\% for sufficiently slow magnetic field ramps.
Here, the efficiency is primarily limited by the preparation of atom pairs in the ground state of relative motion before magnetoassociation.
For typical ramp speeds used in the rest of this work, we expect that over 99\% of atom pairs in the ground state of relative motion are transferred adiabatically to the state \feshbach.

\subsection{Transfer to molecular ground state}
We transfer molecules in state \feshbach to the rovibrational ground state of the $\mathrm{X}^{1}\Sigma^{+}$ potential using a two-photon stimulated Raman adiabatic passage (STIRAP) process \cite{Bergmann1998, Vitanov2017}. 
We populate the hyperfine ground state $\ket{G} \equiv \ket{N=0,M_N=0,m_\mathrm{Rb}=3/2,m_\mathrm{Cs}=7/2}$ at a magnetic field of 181.699(1)\,G~\cite{Takekoshi2014,Molony2014,Gregory2015,Molony2016,Guttridge2023}.
Here, $N$ is the rotational angular momentum quantum number and $M_N$ is its projection along the quantization axis.
$m_{\mathrm{Rb}}$ ($m_{\mathrm{Cs}}$) is the projection of the nuclear spin $i_{\mathrm{Rb}}= 3/2$ ($i_{\mathrm{Cs}}= 7/2$) of Rb (Cs) along the quantization axis.
The state \ground is the lowest energy hyperfine state for magnetic fields above $\sim90$\,G and conveniently, being a spin-stretched state, has well defined nuclear spin projections~\cite{Gregory2016}.

The STIRAP process uses the ``pump'' beam ($\lambda = 1557$\,nm) and a ``Stokes'' beam ($\lambda = 977$\,nm) that together couple state \feshbach to state \ground via the electronically excited state \intermediate~\cite{Takekoshi2014,Molony2014}, as shown in Fig.~\ref{fig:molecule_formation}(a).
The intensities of these beams are modulated with the so-called ``counter-intuitive'' pulse sequence \cite{Bergmann1998} over 70\,$\mu$s, during which the dark state adiabatically evolves from state \feshbach to state \ground. Further details of the STIRAP setup are given in Appendix~\ref{app:experiment}.

\subsection{Molecule detection scheme} \label{sec:mol_det_scheme}
The lack of closed optical cycling transitions in RbCs precludes scattering enough photons for single molecule fluorescence detection. Instead we map the success or failure of molecule formation onto atoms in specific tweezers and then use standard atomic fluorescence imaging of Rb and Cs, as described below.

We utilize a technique similar to erasure conversion in neutral-atom arrays \cite{Scholl2023} to detect sites of the array in which molecule formation failed. Formation errors result in atom pairs remaining in the 1066\,nm tweezers after the molecules have been transferred to state \ground, as shown in Fig.~\ref{fig:overview}(b)(ii). We detect these errors by pulling out the remaining Rb atoms and storing them in a separate row of 817\,nm tweezers (the ``detection array''), as shown in Fig.~\ref{fig:overview}(b)(iii). In addition, we apply resonant light to remove any remaining Cs atoms.

Subsequently, we also reverse the STIRAP sequence to transfer molecules back to state \feshbach before immediately reversing the association field ramps to convert the molecules back to atom pairs.
The resulting atom pairs are then separated by pulling out the Rb atoms and returning them to their original traps, as shown in Fig.~\ref{fig:overview}(b)(iv).

Finally, at the end of the experimental run we take a fluorescence image of Rb and Cs to determine the occupancy of the \emph{three} tweezer arrays: the original arrays containing atoms recovered from the molecules \emph{and} the detection array containing Rb atoms in sites where molecule formation failed. 

From the final fluorescence image we determine the recovery probability $P_r$ of the molecules as follows. Firstly, the presence of a Rb atom in the detection array indicates that molecule formation failed in the corresponding 1066\,nm trap and we ignore that site when analyzing statistics. Conversely, if the detection trap is empty, we assume that a molecule was formed in that site and therefore consider the occupancy of the corresponding initial Rb and Cs traps.
A molecule is then deemed to be `recovered' if both atoms that formed it are successfully imaged in their original traps at the end of a routine.
Thus $P_r$ is defined as the probability that we recover both a Rb and a Cs atom in their initial traps, ignoring sites in which the presence of a Rb atom in the detection array indicates that molecule formation failed.

\begin{figure}
\includegraphics[width=\hsize]{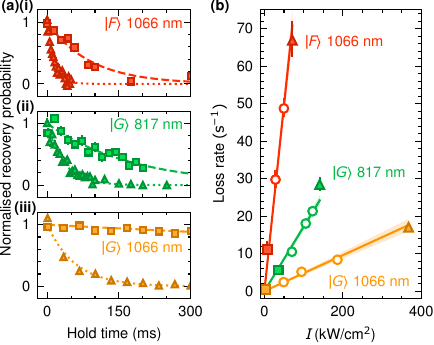}
\caption{Lifetime of RbCs molecules in optical tweezers with wavelengths of 1066\,nm and 817\,nm. (a) 
Normalized recovery probabilities as a function of the hold time in the tweezers for molecules (i) in state \feshbach and 1066\,nm, (ii) the ground state \ground and 817\,nm and (iii) the ground state \ground and 1066\,nm. In each panel results are shown for two different intensities with the values indicated by the corresponding symbols in (b). The axes have been rescaled to make the contrast of the fits equal to unity. The gold squares in (iii) show the loss of molecules in state \ground at our typical operating intensity. (b)~Scaling of the molecule loss rates with tweezer intensity ($I$). The solid lines show linear fits to the measured loss rates.}
\label{fig:lifetimes}
\end{figure}

\subsection{Molecule formation and detection efficiencies} \label{sec:mol_det_and_form_eff}

The efficiency of molecule formation (and subsequent recovery) is primarily limited by the STIRAP transfer efficiency and the loss of molecules in state \feshbach. We quantify these losses below using the detection scheme described in Sec.~\ref{sec:mol_det_scheme} above.

We measure the one-way STIRAP efficiency by repeating many round trips $\ket{F}\rightarrow\ket{G}\rightarrow\ket{F}$ before reversing the association field ramps and measuring the molecule recovery probability. The results are shown in Fig.~\ref{fig:molecule_formation}(d). From this measurement we extract a one-way transfer efficiency of 96.4(1)\%, assuming the efficiency of the forward ($\ket{F}\rightarrow\ket{G}$) and reverse ($\ket{G}\rightarrow\ket{F}$) transfers to be the same. This is marginally better than the efficiencies reported for RbCs in bulk gases \cite{Takekoshi2014,Molony2016} and comparable to the highest reported efficiencies for ground-state transfer of diatomic molecules \cite{he2023,christakis2023}.

Figure~\ref{fig:lifetimes} shows lifetime measurements of the molecular states \feshbach and \ground for different tweezer intensities. 
We find that molecules in the weakly-bound state \feshbach exhibit a much larger loss rate than ground-state molecules.
We have previously observed a photoassociation resonance from state \feshbach at $1063.91(7)$\,nm with an estimated transition dipole moment (TDM) of $0.064(2)\times ea_0$ \cite{thesis:spence}.
We believe that the tweezer light is driving a transition from the $a^{3}\Sigma^{+}$ manifold to the $c^{3}\Sigma^{+}$ manifold~\cite{delieucomm}. 
To reduce the loss rate from photon scattering in the wings of this resonance, we operate the molecule tweezers at a wavelength of $1065.512$\,nm. At this wavelength, we determine the loss rate of molecules in state \feshbach to be {0.99(4)\,s$^{-1}$/(kW/cm$^{2}$)} from the fit to the red points in Fig.~\ref{fig:lifetimes}(b).
To mitigate this loss, we operate at low tweezer intensities and minimize the time between the molecule entering state \feshbach and being transferred to state \ground. Unfortunately, the narrow ($\sim$ 100\,kHz) linewidths of the STIRAP transitions necessitate a 10\,ms hold following the magnetoassociation ramps to achieve sufficient magnetic field stability ($\sim 50$\,mG) for efficient transfer. During this time the molecules in state \feshbach are held in tweezers with an intensity of 6\,kW/cm$^{2}$ such that 5(1)\% are lost.
We note that the molecular state \s shown in Fig.~\ref{fig:molecule_formation}(b) that is populated in the initial magnetoassociation ramp has a much longer lifetime, consistent with that of the atom pair ($>10$\,s). However, STIRAP from this state is inefficient due to the weak coupling to state \intermediate.

The loss rate of molecules in the ground state \ground is much lower than that of molecules in state \feshbach. From the fits in Fig.~\ref{fig:lifetimes} (b), we determine loss rates for state \ground of {0.171(8) s$^{-1}$/(kW/cm$^{2}$)} in tweezers with a wavelength of 816.848\,nm and {0.047(4) s$^{-1}$/(kW/cm$^{2}$)} for a wavelength of 1065.512\,nm. The linear relation observed between loss rate and intensity suggests that the lifetime is limited by photon scattering of the tweezer light, most likely Raman scattering. A single Raman scattering event would appear as loss since we only detect molecules in the specific rotational and hyperfine state addressed by the STIRAP lasers. In light of this, we typically operate the tweezers at a low intensity where the lifetime of state \ground is typically 2.7(4)\,s, corresponding to the gold squares in Fig.~\ref{fig:lifetimes}(a)(iii), and loss is negligible for the duration of most experiments. 
This lifetime is still limited by scattering of the trapping light; the vacuum lifetime of atoms in the experiment is $>30$\,s~\cite{thesis:brooks}.

\begin{table}
    \centering
       \begin{tabular}{l|c}
        \textbf{Experimental stage} & \textbf{Efficiency} \\
        \hline
        \hline
        Preparation of atom pair hyperfine state  & 0.93(2) \\
        Occupancy of ground state of relative motion\\\hspace{10mm}(following the merging of the traps) & 0.56(5)\\
        Magnetoassociation efficiency (calculated) & $> 0.99$ \\
        \hline
        \textbf{Atom pair $\rightarrow$ \feshbach conversion} & \textbf{0.53(1)} \\
        \hline
        \hline
        \feshbach survival pre-STIRAP & 0.95(1) \\
        STIRAP transfer $\ket{F} \rightarrow \ket{G}$ & 0.964(1)\\
        \hline
        \textbf{\feshbach $\rightarrow$ \ground conversion (calculated)} & \textbf{0.91(1)}\\
        \hline
        \hline
        \textbf{Atom pair $\rightarrow$ \ground conversion (calculated)} & \textbf{0.48(2)}
    \end{tabular}
\caption{Efficiencies of each stage of the molecule formation. The values are measured experimentally unless stated otherwise.}
    \label{tab:mol_formation_eff}
\end{table}

The efficiencies of each step of the molecule formation protocol are summarized in Table~\ref{tab:mol_formation_eff}.
We successfully convert atom pairs to molecules in state \feshbach with an efficiency of 53(1)\%, limited by the initial state preparation of the atom pairs.
Subsequently, 91(1)\% of the molecules in state \feshbach are successfully transferred to the ground state \ground.
The overall efficiency for the conversion of an atom pair to a rovibrational ground state molecule is therefore 48(2)\%.

The maximum probability of molecule recovery that we measure is $P_r = 75(1)\%$. This corresponds to the value after a single round-trip STIRAP in Fig.~\ref{fig:molecule_formation}(d). If our scheme for detecting molecule formation was perfect, we would expect to measure 88(1)\%, limited by the lifetime of state \feshbach and the infidelity of a round-trip STIRAP.
However, our detection scheme overestimates molecule formation due to loss of atom pairs prior to magnetoassociation. The overall probability that a detection trap is empty is 60(3)\%; a combination of successful molecule formation (53(1)\%) and loss of atoms prepared in the wrong hyperfine state (7(2)\%). 
With our detection scheme, these two events are indistinguishable and we assume that a molecule has been formed in both cases. In reality, a molecule is only formed in 88(3)\% of cases where the detection trap is empty.
Accounting for this, we would expect to measure $P_r = 77(3)\%$, in agreement with our observations.

\begin{figure*}
\includegraphics[width=0.95\hsize]{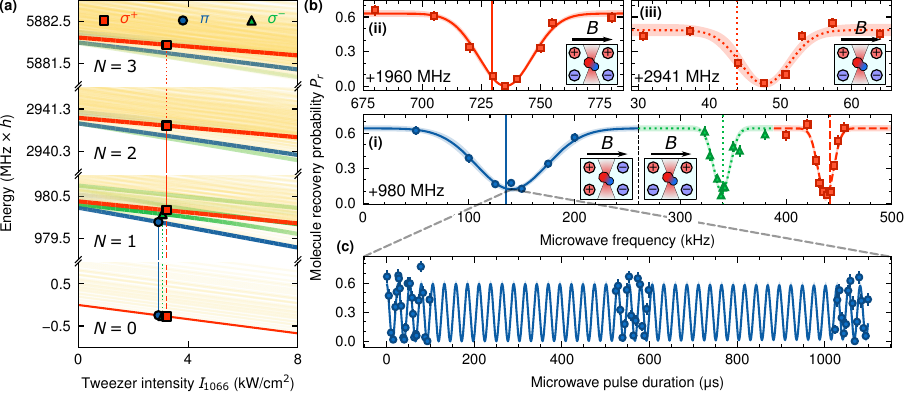}
\caption{Excitation of RbCs molecules to higher rotational states. (a) Hyperfine energy structure of the first four rotational manifolds of RbCs as a function of 1066\,nm tweezer intensity at a magnetic field of 181.699\,G. Energies are given relative to the energy of state \ground in free space. The red lines show the spin-stretched hyperfine states of each manifold, where the molecule is initially prepared in the ground state (lowest red line). \sigmap, $\pi$, and \sigmam transitions from each spin-stretched state are shown in red, blue, and green respectively. The yellow lines show other hyperfine states that we do not access. The color of the lines represents the TDM of a given transition: more intense lines have higher TDMs. (b) Spectroscopy of the $N=1,2,3$ rotational manifolds for (i),(ii),(iii) respectively in a 3.07 kW/cm$^2$ tweezer. Changing the electrode configuration (insets) allows us to drive either $\pi$ or \sigmapm transitions. (c) Rabi oscillation on the $\pi$ transition $\ket{0,0}\rightarrow\ket{1,0}$ for a single trapped molecule.}
\label{fig:mw_spec}
\end{figure*}

\section{Global Control of Rotational States}\label{sec:global}
The lowest rotational energy levels in the vibrational ground state of RbCs are shown in Fig.~\ref{fig:overview}(c) where states are labeled with $\ket{N,M_N}$.
The energies of the rotational levels are $h\times B_{\nu} N(N+1)$ such that the splitting between neighboring rotational manifolds is in the microwave domain (for RbCs, $B_{\nu} \approx 490$\,$\mathrm{MHz}$) \cite{Gregory2016}.
This picture ignores coupling between the rotational angular momentum and the nuclear spins of the constituent atoms ($i_\mathrm{Rb} = 3/2$ and $i_\mathrm{Cs} = 7/2$).
When optical and magnetic fields are applied, this coupling causes each $\ket{N,M_N}$ state to split into $(2i_\mathrm{Rb}+1)(2i_\mathrm{Cs}+1) = 32$ hyperfine states. 
This hyperfine structure is illustrated in Fig.~\ref{fig:mw_spec}(a) for the lowest four rotational states.
Here, the energies of the states are shown as a function of the intensity $I_{1066}$ of the 1066\,nm tweezer and the magnetic field is 181.699\,G. 
When performing STIRAP we form molecules in the rovibrational and hyperfine ground state \ground (lowest red line).
The transitions that we drive between rotational levels are shown by vertical lines.

Transitions between rotational levels are driven using microwave fields to which strong coupling is facilitated by the molecule-frame electric dipole moment (for RbCs, $d_0 =  1.225$\,D) \cite{Molony2014}.
Allowed electric dipole transitions are those with $|\Delta N| = 1$ and $|\Delta M_N| \le 1$. The strength of the transition is determined by the transition dipole moment (TDM) $\mathbf{\mu}_{i,j}=\bra{\psi_{i}}\mathbf{\mu}\ket{\psi_{j}}$, where the components $(\mu^{z}_{i,j} ,\mu^{+}_{i,j} ,\mu^{-}_{i,j})$ of $\mathbf{\mu}_{i,j}$ describe the strength of $\pi$, \sigmap and \sigmam transitions respectively.
The nuclear spin is not addressed when driving rotational transitions such that we can only couple to hyperfine states with nuclear spins unchanged from those of \ground; namely $m_\mathrm{Rb}=3/2$ and $m_\mathrm{Cs}=7/2$.
At the operating fields used in this work ($\sim 200$\,G) the molecular eigenstates are generally superpositions of states of different $m_\mathrm{Rb}$ and $m_\mathrm{Cs}$, and the only good quantum numbers that can be used to describe them are $N$ and $M_F \equiv M_N + m_\mathrm{Rb}+m_\mathrm{Cs}$.
The exceptions to this are the stretched states with maximum $\left|M_F\right|$; for these states $\left|m_{\rm{Rb}}\right| = 3/2$, $\left|m_{\rm{Cs}}\right| = 7/2$ and $M_N$ is a good quantum number.
For the work presented here, we drive transitions to either stretched states or hyperfine states with mixed character for which the component with $m_\mathrm{Rb}=3/2$ and $m_\mathrm{Cs}=7/2$ has the largest amplitude.
This criterion selects the transitions with the highest TDMs.
For simplicity, we continue to label the states $\ket{N,M_N}$, but give the full state compositions in Appendix~\ref{app:mol_states}.

We use in-vacuum electrodes mounted inside the glass cell \cite{Brooks21} as a microwave antenna to drive coherent transfer between molecular rotational states. These electrodes were designed to orient molecules in the laboratory frame by generating large dc electric fields.
The four electrodes are positioned in a 9.6\,mm $\times$ 5.6\,mm rectangular array centered around the optical tweezers: this aspect ratio ($\sqrt{3}:1$) increases the uniformity of applied fields by eliminating the field curvature along the horizontal axis~\cite{Gempel2016}.
We find that the electrodes are a good antenna for the $\sim$\,GHz frequency radiation that is resonant with RbCs rotational transitions.
The magnetic field which sets the quantization axis is applied in the horizontal direction parallel to the long dimension of the rectangular electrode array.
An additional external dipole Wi-Fi antenna is mounted approximately 10\,cm from the vacuum chamber. 
Using this external antenna we can also drive transitions, albeit with much reduced polarization control due to the presence of magnetic field coils around the cell.

We demonstrate rotational state control by driving coherent microwave transitions from the rovibrational ground state \ground to higher rotational states.
Excitation to higher rotational states is detected by the failure to recover atom pairs from the (excited) molecules at the end of the experimental sequence due to the state specificity of the reverse STIRAP transfer.
We selectively drive either \sigmapm or $\pi$ transitions by connecting the electrodes in different configurations to change the orientation of the electric field of the microwave radiation.
When the electric field is parallel to the applied magnetic field we drive $\pi$ transitions; when the two fields are orthogonal we drive \sigmapm transitions.

Figure~\ref{fig:mw_spec}(b)(i) shows spectroscopy from state \ground to the $N=1$ manifold, with the polarity of the connections to the electrodes shown inset. Here, the magnetic field is 181.699(1)\,G and the intensity of the 1066\,nm tweezer is $I_{1066} = 3.07$\,kW/cm$^{2}$.
We measure the frequencies of the $\pi$, \sigmam, and \sigmap transitions to be 980.140(2)\,MHz, 980.3391(9)\,MHz and 980.4374(5)\,MHz, respectively.
The widths of the measured features are transform-limited.
The vertical lines in Fig.~\ref{fig:mw_spec}(b) show the expected transition frequencies. 
We calculate these and the state energies shown in Fig.~\ref{fig:mw_spec}(a) by solving the molecular Hamiltonian \cite{blackmore2023}, including the interactions with external optical and magnetic fields. We use the molecular constants determined in previous bulk-gas experiments~\cite{Gregory2016,blackmore2020b,gregory2021}.
The value of the isotropic polarizability $\alpha^{(0)}$ is scaled from that measured by Blackmore \emph{et al.} \cite{blackmore2020} to account for the difference in trapping wavelengths \cite{Vexiau2017}; here we use $\alpha^{(0)}_{1066} = 2000\times 4\pi\varepsilon_0 a_0^3$.
By measuring the frequency of the $\pi$ transition in 1065.512\,nm tweezers of different intensities $I_{1066}$, we determine the anisotropic polarizability $\alpha^{(2)}_{1066} = 1980(60) \times 4\pi\varepsilon_0 a_0^3$ \cite{preparation}.
The measured transition frequencies are within 10\,kHz of the calculated values; we expect that the discrepancy between the two is primarily caused by our simplifying assumption that the polarization of the tweezer is exactly aligned to the quantization axis of the magnetic field.

We observe coherent oscillations between states by changing the duration of the applied microwave pulses.
For example, in Fig.~\ref{fig:mw_spec}(c) we show the effect of changing the pulse length with the microwave frequency set to that of the $\pi$ transition $\ket{0,0}\rightarrow\ket{1,0}$ for a single trapped molecule.
With a small RF power of -16\,dBm incident to the electrodes, we obtain a Rabi frequency of 37.96(2)\,kHz  and observe no significant damping in the contrast after approximately 40 Rabi oscillations. 
The microwave field produced by the electrode array is highly linearly polarized.
For example, with the field set to drive $\pi$ transitions, we are not able to resonantly drive the \sigmam transition, even when the Rabi frequency on the $\pi$ transition is increased to 133.7(1)\,kHz.
Setting a conservative upper bound on Rabi frequency with which we drive the \sigmam transition of 1\,kHz, we extract the linear polarization purity of the microwave radiation emitted by the electrode array to be in excess of $10^{4}:1$. This enables high-fidelity control of the rotational states.

We probe higher rotational manifolds in the molecule using successive microwave transitions \cite{blackmore2020b}.
Here, we restrict ourselves to \sigmap transitions so that we always occupy a stretched hyperfine state in each rotational manifold.
For example, in Fig.~\ref{fig:mw_spec}(a)(ii) we present spectroscopy of the transition $\ket{1,1}\rightarrow\ket{2,2}$.
This is measured by first performing a $\pi$ pulse on the transition $\ket{0,0}\rightarrow\ket{1,1}$ to prepare the molecule in $\ket{1,1}$ prior to the spectroscopy pulse.
After the spectroscopy pulse, a third microwave pulse returns any molecules remaining in $\ket{1,1}$ back to $\ket{0,0}$ from which atom pairs can be recovered.
Molecules that were excited to $\ket{2,2}$ during the spectroscopy pulse are not returned back to $\ket{0,0}$, resulting in atom pairs not being recovered.
This is easily extended to higher manifolds; we generally prepare molecules in the stretched hyperfine state $\ket{N,N}$ with a series of $N$ coherent $\pi$ pulses before probing the transition $\ket{N,N}\rightarrow\ket{N+1,N+1}$ and returning molecules in $\ket{N,N}$ to $\ket{0,0}$.
For example, in Fig.~\ref{fig:mw_spec}(a)(iii), we perform similar spectroscopy of the transition $\ket{2,2}\rightarrow\ket{3,3}$ with this procedure. 
As before, the measured frequencies for these transitions are within 10\,kHz of the predicted frequencies indicated by the vertical lines in Fig.~\ref{fig:mw_spec}(b).
Extension to more rotational states will allow the realization of a large number of synthetic lattice sites with fully controllable synthetic inter-site tunnelings for engineering synthetic band structures \cite{Sundar2018}.

\section{Multi-state readout}\label{sec:readout}
Proposed quantum simulators composed of molecules often utilize the rotational states to encode  pseudo-spins~\cite{Micheli2006,Gorshkov2011}. The detection of multiple rotational states of a molecule in a single iteration of the experiment would therefore prove highly valuable, particularly given the finite efficiency of forming bialkali molecules. For example, without the ability to readout multiple molecular states, it is impossible to distinguish between a molecule which is lost and a molecule in a spin state which is not detected. In the following we describe a technique that can be used to unambiguously detect the rotational state of a molecule on a given site. 

\begin{figure}
\includegraphics[width=\hsize]{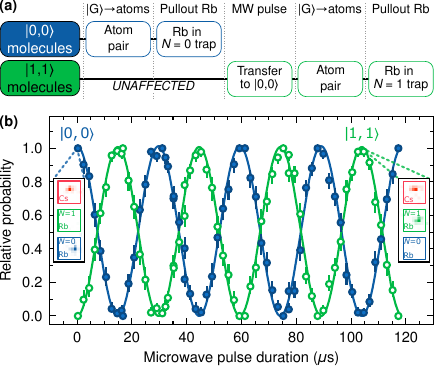}
\caption{Readout of multiple rotational states whilst driving the transition $\ket{0,0}\rightarrow\ket{1,1}$. (a) Flowchart of the detection procedure as described in the text. (b) Coherent transfer between $\ket{0,0}$ and $\ket{1,1}$. The Rabi frequency is 33.69(3)\,kHz and the fitted contrast is consistent with unity. The insets show the Cs (red) and Rb (green and blue) traps with example fluorescence images obtained from a molecule in states $\ket{0,0}$ (left) and $\ket{1,1}$ (right). Each point corresponds to the average across the four molecule array with 100 experimental repetitions, where 24\% of these runs satisfy the post-selection criteria.}
\label{fig:state_sensitive_detection}
\end{figure}

We present an experimental scheme that maps the rotational state of the molecule onto atoms in spatially distinct tweezers, similar to the proposal of Ref.~\cite{Covey2018a} where the state is mapped onto the atomic species. In our scheme, we detect the internal state of the molecule by mapping it onto the position of a Rb atom in the final fluorescence image.
A flowchart of the detection scheme is shown in Fig.~\ref{fig:state_sensitive_detection}(a).
We exploit the state specificity of the reverse STIRAP transfer; only molecules in state \ground are converted into atom pairs during the reverse STIRAP pulses and dissociation magnetic field sweeps.
Molecules in excited rotational states are unaffected by these stages of the experimental routine.
After atom pairs are recovered from molecules that were in state \ground, they are separated and the Rb atoms are stored in a row of 817\,nm tweezers.
We then return to the usual operating magnetic field to transfer molecules in excited rotational states back to state \ground with a series of microwave pulses and repeat the dissociation steps.
However, this time when separating the resultant atom pairs, we place the Rb atoms in a \emph{different} row of 817\,nm tweezers.
After all the molecules have been dissociated into atom pairs, the magnetic field is reduced to 4.78\,G and a final fluorescence image is taken. With this image, we can detect the rotational state of the molecule prior to the readout procedure by observing which tweezer the Rb atoms populate.
Using mid-sequence detection of Rb atoms, would allow this procedure to be repeated multiple times. This would enable many internal molecular states to be readout in a single experimental run, ideal for implementations of qudits~\cite{Sawant2020} or quantum error correction using the internal states of the molecule~\cite{Albert2020}.

This detection scheme allows us to mitigate the effects of noise in our experimental data. Such noise can result from fluctuations in the molecule formation efficiency or molecule loss which reduces the recovery probability, $P_r$.
We are also able to eliminate leakage errors that occur when molecules leave a chosen set of energy levels. The lowest rotational levels of RbCs have lifetimes exceeding 1000\,s limited by black-body radiation~\cite{Kotochigova2005}. Consequently, leakage errors due to off-resonant excitation during microwave transfers, for example, are much more likely than bit-flip errors for RbCs qubits. Using this detection scheme, we specify the energy level subspace that we wish to readout with the choice of microwave pulses prior to converting the molecules back to atom pairs.

In Fig.~\ref{fig:state_sensitive_detection}(b) we present an example measurement performed with this detection scheme.
Here, we drive a Rabi oscillation on the \sigmap transition between $\ket{0,0}$ (blue filled circles) and $\ket{1,1}$ (green empty circles) with a resonant microwave pulse applied to an array of four molecules.
The Rabi frequency with which we drive the transition is 33.69(3)\,kHz. This avoids significant off-resonant excitation of the \sigmam transition which is detuned by $-96(1)$\,kHz at the magnetic field used in the experiment, as shown in Fig.~\ref{fig:mw_spec}(b)(i). 
After this pulse, molecules in $\ket{0,0}$ are converted back to atom pairs from which Rb atoms are moved to the ``$N=0$ detection'' traps (insets, blue square).
Molecules in $\ket{1,1}$ are then transferred back to $\ket{0,0}$ with a $\pi$ pulse before we convert them back to atom pairs and deposit the Rb atoms into the ``$N=1$ detection'' traps (insets, green square).
Cs atoms always remain in the 1066\,nm traps in which the molecules are formed (insets, red square).
We post-select data to consider only experimental runs in which both a Cs atom and a Rb atom (in either of the two detection traps) are successfully recovered from an initial atom pair.
This corresponds to 24\% of the total number of runs for this dataset.
The relative occupation of Rb atoms in the detection traps is used to infer the state of the molecule before the detection procedure.
The fitted contrast of the Rabi oscillations is consistent with unity and we do not observe dephasing over the range of pulse durations shown here. 

\section{Local Control of Rotational States}\label{sec:local}
Controlling the rotational states of individual molecules within an array is essential for a range of applications. For example, preparing  reactants in distinct rotational states facilitates studies of state-controlled quantum chemistry~\cite{Liu2021b}. Additionally, certain quantum computation architectures using ultracold molecules require the selective excitation of molecules to perform single qubit gates~\cite{DeMille2002} or to execute entangling gates between chosen pairs of molecules using microwave fields~\cite{Hughes2020}. The targeted transfer of subsets of molecules into non-interacting states allows them to be shelved for mid-circuit readout, enabling measurement-based quantum computation~\cite{Raussendorf2001,Briegel2009} or the study of measurement-induced phase transitions~\cite{Fisher2023}.

\begin{figure}
\includegraphics[width=\hsize]{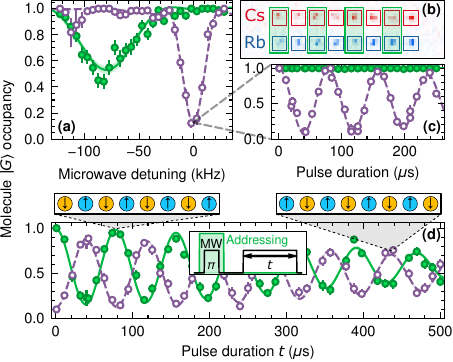}
\caption{Local control of rotational states in an array of molecules. (a) Selected molecules are addressed with an additional 817\,nm tweezer that causes a light shift of the rotational transition in these molecules (green filled points) relative to the unaddressed molecules (purple empty points). (b) An image of the atoms that form the molecular array showing which sites are addressed (green highlight). (c) For a light shift of approximately $-200$\,kHz applied to the addressed molecules, we drive Rabi oscillations in the unaddressed molecules only. (d) Preparation and manipulation of alternating spin chains of molecules. An initial $\pi$ pulse transfers only unaddressed molecules to $\ket{\uparrow}$. The addressing is then removed to coherently rotate the spins of all molecules in the array.}
\label{fig:addressing}
\end{figure}

We demonstrate site-resolved control of the rotational state using an additional array of optical tweezers to address selected molecules.
The additional tweezers cause a differential light shift between molecular states, altering the microwave transition frequency on the addressed sites.
An example of this is shown in Fig.~\ref{fig:addressing}(a).
In this measurement every other trap in an eight-trap array is addressed with an additional 817\,nm tweezer, as indicated by the green rectangles in Fig.~\ref{fig:addressing}(b). The addressing tweezers are ramped up to an intensity of $2.18$\,kW/cm$^2$ after the molecules have been prepared in state \ground. We then perform microwave spectroscopy on the array. Following this, the addressing light is removed such that all molecules are resonant with the microwave $\pi$ pulses required for the multi-state readout described earlier and the reverse STIRAP. The results in Fig.~\ref{fig:addressing}(a) show that the additional 817\,nm tweezer light causes the frequency of the $\ket{0,0} \rightarrow \ket{1,1}$ transition to shift by $-80(2)$\,kHz in the addressed molecules (green filled circles) relative to the unaddressed molecules (purple empty circles).
The observed light shifts of the rotational transitions allow us to extract a value for the anisotropic polarisability of $\alpha^{(2)}_{817} = -2814(12) \times 4\pi\varepsilon_0 a_0^3$ for the 817\,nm addressing tweezer \cite{preparation}. We note the increase in the size of the error bars for unaddressed molecules at a detuning of around $-100$\,kHz in Fig.~\ref{fig:addressing}(a). This results from these molecules being excited on the \sigmam transition such that the number of molecules remaining in the $\{\ket{0,0},\ket{1,1}\}$ subspace from which we sample is greatly reduced.

When the induced light shift is much larger than the Rabi frequency of our chosen transition, we are able to drive transitions in \emph{only} the unaddressed molecules.
We demonstrate this in Fig.~\ref{fig:addressing}(c) where we increase the intensity of the addressing tweezers such that the light shift of the transition is approximately $-200$\,kHz.
We then apply microwave radiation that is resonant for the unaddressed molecules to drive a Rabi oscillation in only these molecules.
After a $\pi$ pulse, we observe no transfer of the addressed molecules out of state \ground and place a $1\sigma$ upper bound of 1.0\% on the probability of driving an undesired excitation.
We chose to target unaddressed molecules as fluctuations in the relative alignment of the trapping and addressing tweezers cause variations in the induced light shift and broaden the transitions of the addressed molecules.

We dynamically switch the addressing on and off during the experimental sequence to change between driving molecular transitions locally and globally.
As an example, Fig.~\ref{fig:addressing}(d) shows the result of an experiment where we form an alternating spin chain of molecules with $\ket{\downarrow}\equiv\ket{0,0}$ and $\ket{\uparrow}\equiv\ket{1,1}$ and then drive Rabi oscillations in the whole array. 
The molecule formation stages initialize the array in $\ket{\downarrow}$.
As before, half of the molecules in the array are then addressed with 817\,nm light.
A $\pi$ pulse on the $\ket{\downarrow} \rightarrow \ket{\uparrow}$ transition is then driven in only the unaddressed molecules to prepare an alternating spin chain.
We then remove the addressing light such that a second microwave pulse drives the rotational transition for all molecules in the array.
This pulse rotates all the spins in the chain such that two adjacent molecules are always out of phase with each other.
The dephasing evident in Fig.~\ref{fig:addressing}(d) is primarily caused by different trap depths across the array of 8 molecules. This leads to a variation in the differential light shifts along the spin chain, such that the microwave field is not exactly resonant with all the molecules. In future work, we plan to address this problem by using an array of tweezers at a magic wavelength such that the differential light shift between the states $\ket{\downarrow}$ and $\ket{\uparrow}$ is eliminated~\cite{gregory2024}.

\section{Deterministic Array Preparation}\label{sec:rearrangement}
\begin{figure}
\includegraphics[width=0.95\hsize]{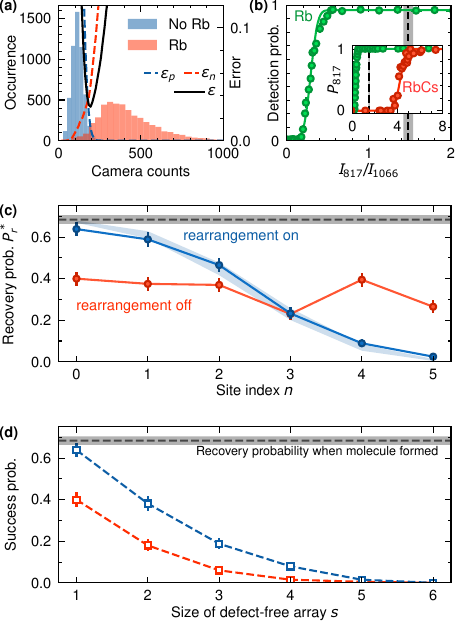}
\caption{Rearrangement of assembled molecules. (a) Histogram of camera counts obtained using high-field imaging of Rb atoms. Red (blue) data are counts obtained when an atom is (is not) present. The blue and red dashed lines are the probabilities of false-positive $(\varepsilon_p)$ and false-negative $(\varepsilon_n)$ errors, respectively, as the occupancy threshold is changed. The black solid line is the average error $(\varepsilon)$. (b) The probability of detecting Rb in the detection trap as a function of the intensity ratio between the 817\,nm and 1066\,nm tweezers, $I_{817}/I_{1066}$, during the Rb pullout. The vertical dashed line indicates the value used in a typical sequence. The inset shows the probability $P_{817}$ of pulling out Rb (green) and RbCs (red) with the 817\,nm tweezer for higher values of $I_{817}/I_{1066}$. (c) The probability of molecule recovery $P_r^*$ for each site $n$ of the array. Blue (red) points show data with (without) molecule rearrangement for an initial array of six atom pairs. No post-selection on successful molecule formation is performed; the black dashed line shows the measured recovery probability ($0.68(2)$) for a molecule that is formed and left in a single trap. The blue shaded region shows the prediction of a Monte Carlo simulation of the rearrangement. 
(d) The probability of successfully recovering defect-free arrays of size $s$, starting from site zero, with (blue) and without (red) rearrangement.}
\label{fig:rearrangement}
\end{figure}

We now demonstrate the preparation of defect-free arrays of molecules.
The primary source of configurational entropy in our array is the finite conversion efficiency of atom pairs to molecules which leads to some tweezers not containing molecules.
We remove this entropy by detecting the traps where molecule formation failed using the procedure described in Sec.~\ref{sec:mol_det_scheme}.
Unlike the experiments presented thus far, in this experiment we perform the detection mid-sequence and use the information to rearrange molecules to occupy sites where formation failed.

Mid-sequence detection of molecule formation errors requires imaging Rb atoms at the magnetic field of $181.699(1)$\,G used for STIRAP. At the normal imaging field of 4.78\,G, the state \ground is no longer the lowest in energy and the hyperfine levels are more closely spaced. Imaging at high magnetic field therefore avoids potential loss of molecules to other states due to sweeping the magnetic field through numerous level crossings. 
Rb atoms in the detection array are imaged on the closed transition ${(5\mathrm{s}_{1/2},f=2,m_f=2)} \rightarrow {(5\mathrm{p}_{3/2}, f'=3,m'_f=3)}$  (hereafter ${(2,2)}$ and ${(3',3')}$, respectively). This approach has previously been used for non-destructive hyperfine-state readout of individually trapped Rb atoms \cite{Kwon2017,MartinezDorantes2017}.
As the Rb atoms are initially in the state ${(5\mathrm{s}_{1/2}, f=1,m_f=1)}$ required for molecule formation, they are transferred to the state ${(2,2)}$ with microwave adiabatic rapid passage (ARP) before imaging, as illustrated in Fig.~\ref{fig:overview}(d). Further details of the detection scheme are given in Appendix~\ref{app:experiment}.

Figure~\ref{fig:rearrangement}(a) shows a histogram of camera counts from a single Rb trap obtained using the high-field imaging procedure.
During a rearrangement routine, trap occupancy is determined by comparing whether the observed counts are above or below a predefined threshold.
The lines in Fig.~\ref{fig:rearrangement}(a) show the error in the occupancy assignment as this threshold is changed; the blue dashed line is the false positive error $\varepsilon_p$ and the red dashed line is the false negative error $\varepsilon_n$. The black solid line is the average error probability $\varepsilon$, from which we extract a value of 3\% when the threshold is optimized.

In Fig.~\ref{fig:rearrangement}(b) we verify the performance of the detection scheme by varying the ratio of tweezer intensities $I_{817}/I_{1066}$ during the Rb pullout step. 
When $I_{817}$ is too low, no Rb atoms are moved into the detection tweezers and a non-zero probability of detection corresponds to a false positive. 
Conversely, when $I_{817}$ is high, all remaining Rb atoms are transferred to the detection tweezer and a probability below unity corresponds to a false negative. 
From the fit to Fig.~\ref{fig:rearrangement}(b) we find the combined procedure of pullout and imaging gives a false positive rate of 0.7(1)\% and a false negative rate of 3.6(1)\%. 
The latter is dominated by a $\sim 2$\% probability for loss of the Rb atom prior to imaging. This value is consistent with the trap lifetime of Rb atoms in the experiment ($\sim 30$\,s) and the duration of a typical experimental routine after Rb has been loaded ($\sim500$\,ms).
The vertical dashed line in Fig.~\ref{fig:rearrangement}(b) shows the intensity ratio of 1.48(6) used for mid-sequence detection. This value is chosen to saturate the Rb detection fidelity whilst leaving molecules in their original traps, as shown in the inset.

We use the real-time information obtained from the high-field image to identify traps in which molecule formation was successful and rearrange the molecules to one side of the array. 
Molecule occupancy is assigned by inverting the measured Rb occupancy in the corresponding traps of the detection array.
Molecules are then transferred from the 1066\,nm array to an overlapping 817\,nm array and unoccupied molecule traps are extinguished.
Occupied molecule traps are then shuttled to one end of the array before the molecules are transferred back into the 1066\,nm array.

We show the molecule recovery probabilities $P_r^*$ obtained using this rearrangement scheme in Fig.~\ref{fig:rearrangement}(c). Here, we do \emph{not} post-select statistics based on successful molecule formation, unlike in the experiments presented in earlier sections.
For these measurements, exactly six atom pairs are prepared in the 1066\,nm array which we attempt to associate into molecules and transfer to state \ground.
For points with rearrangement enabled (blue), molecules are shuttled to the end of the array; for points with rearrangement disabled (red), the molecules are left in their original traps.
We then reverse the association routine and image resultant atom pairs to determine $P_r^*$.

With rearrangement disabled, the molecule recovery across the array is approximately uniform with an average of 34(1)\%.
This is consistent with the 36(2)\% that we expect from combining the typical  molecule formation efficiency of 53(1)\% with the molecule recapture probability $P_r = 68(2)\%$ obtained with this experimental routine when post-selecting on successful molecule formation.
We note that the molecule recapture probability is reduced from the values reported earlier in the paper due to increased time spent in the tweezers during imaging and the additional time required for calculating the rearrangement sequence.

When rearrangement is enabled, the average molecule recovery in the array remains 34(1)\% but the distribution is no longer uniform, being weighted significantly towards the low-index sites in the array as intended.
The observed recovery in the array agrees well with the prediction of a Monte Carlo simulation of the rearrangement, indicated by the blue shaded region in Fig.~\ref{fig:rearrangement}(c).
In this simulation, we populate the initial array of molecules by generating a random number, $x_n$, between 0 and 1 for each site in the array. If $x_n$ is lower than the measured molecule formation efficiency for that site, then the site is deemed to be occupied.
Once the initial occupancy of the array is determined, all molecules are shuttled to fill the traps with the lowest site indices.
We assume that no molecules are lost during this process.
This is repeated for 500 initializations of the array, and the average occupancy of each site is determined.
The site occupancies are then scaled by the measured value of $P_r$.
The shaded region shows the $1\sigma$ bounds on the simulation results, obtained by repeating it 500 times using different values of the molecule formation efficiency and $P_r$, both sampled from Gaussian distributions centered about their measured values with standard deviations equal to their experimental uncertainties.

Figure~\ref{fig:rearrangement}(d) shows the probability of successfully observing a defect-free array of size $s$. With rearrangement enabled, this probability scales as $(P_r)^s$ due to the loss of molecules prior to the final fluorescence image.

\section{Scaling to larger arrays}\label{sec:outlook}
\begin{figure}
\includegraphics[width=\hsize]{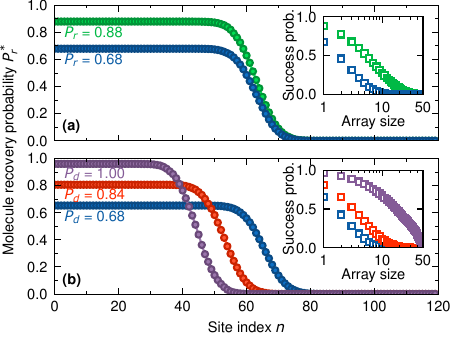}
\caption{Prospects for scaling the molecule rearrangement protocol to larger array sizes. The performance is simulated using the Monte Carlo method described in the text, starting with an array of 120 atom pairs.
The main plots show the probability $P_r^*$ of an atom pair being recovered from a molecule in site index $n$ when \emph{not} post-selecting on molecule formation. The insets show the probability of measuring a defect-free array of a given size. (a) Rearrangement performance for different molecule recapture probabilities $P_r$ using the detection scheme used in this work. (b) Rearrangement performance using direct detection of molecules in state \ground with probability $P_d$, rather than the detection of the failure to form molecules in state \feshbach.}
\label{fig:rearrangement_outlook}
\end{figure}

Finally, we discuss scaling to larger arrays trapping a greater number of molecules. Currently, the performance of our rearrangement protocol is limited by laser power as this determines the initial number of molecules that can be formed. 
Using laser sources that produce 1\,W of 1066\,nm light and 100\,mW of 817\,nm light, we are able to assemble and rearrange an array of molecules starting from six atom pairs. However, at these wavelengths laser sources with output powers of 20\,W and 2\,W, respectively, are readily available. Such lasers would allow a 20-fold increase in the array size in the short term and we note that higher power laser sources exceeding 100\,W \cite{Dixneuf2020,Wellmann2020} are available for further scaling in the long term.

In Fig.~\ref{fig:rearrangement_outlook}(a) we show the results of a Monte Carlo simulation of the expected rearrangement performance when using 120 atom pairs, corresponding to the anticipated 20-fold increase in laser power.
The simulation is the same as that discussed in Sec.~\ref{sec:rearrangement}, but with a larger array of traps.
The blue symbols show the predicted performance for $P_r = 68\%$; the value we measured in the rearrangement routine using six atom pairs.  The green symbols show the performance that would be achieved if the molecule recovery probability were improved to $P_r=88\%$. This latter value requires the infidelity in the hyperfine-state preparation to be reduced to 2\% and the STIRAP efficiency increased to 99\%. Both improvements are feasible in the near-term by changing the wavelength of the Rb tweezer to be further detuned (to reduce Raman scattering) and by suppressing phase noise on the STIRAP lasers using feed-forward techniques~\cite{li2022,chao2023}.
In both cases, the effect of non-unity $P_r$ is to cause false-positive errors when a molecule in state \feshbach is formed but subsequently lost, decreasing the average number of molecules in the array.
The inset in Fig.~\ref{fig:rearrangement_outlook}(a) shows the probability of preparing a defect-free array, which reduces with array size $s$ proportional to $(P_r)^s$, as we observed experimentally in Sec.~\ref{sec:rearrangement}.

We note that increasing the laser power available for tweezer generation will allow an increase in the number of rotational states that can be readout in a single run of the experiment. There is no fundamental limit to the number of states that can be readout with our detection scheme; we need only to have enough laser power to generate the required number of detection arrays.

Developing the capability to non-destructively detect molecules in state \ground would greatly enhance the prospect of large defect-free arrays of RbCs molecules.
Currently, as the assembled molecules cannot be directly imaged, detection is limited to measuring only when a molecule has been formed. Consequently, we cannot correct for the subsequent loss of molecules. This leads to a significant drop in the probability of preparing a defect-free array when scaling to larger systems. To overcome this limitation, we propose to exploit the recently observed long-range interactions between molecules and Rydberg atoms~\cite{Guttridge2023} to detect molecules in state \ground directly.
In such a scheme, atomic Rydberg excitation is blockaded when a ground state molecule is present such that the presence of a molecule can be inferred from the failure to excite to a Rydberg state~\cite{Kuznetsova2016,Zeppenfeld2017,Wang2022,Zhang2022}.

In Fig.~\ref{fig:rearrangement_outlook}(b) we show the expected recovery of atom pairs from molecules in a sorted array using the proposed Rydberg-atom scheme.
Here we use $P_r = 68\%$, the effect of which is now to reduce the average trap occupancy \emph{before} rearrangement as detection can be performed after all the lossy molecule formation stages.
The simulation is similar to before, but a trap is now occupied prior to rearrangement if $x_n < P_f P_r / \mathcal{F}_\mathrm{STIRAP}$, where $P_f=53\%$ is the assumed molecule formation probability and $\mathcal{F}_\mathrm{STIRAP}=96.4\%$ is the STIRAP fidelity.
We show the results of simulations using different values of the molecule detection probability $P_d$.
We expect $P_d$ to be dominated by false-positive errors due to imperfect transfer to the atomic Rydberg state when no molecule is present. We incorporate this into the simulation by assigning each site with a second random number $y_n$ between 0 and 1, such that if $y_n>P_d$ on an unoccupied site, we simulate a false-positive error in the detection by using this trap during the rearrangement even though it is unoccupied.
As before, we repeat this simulation for 500 initializations of the array and determine the average occupancy of each site in the array.
The limiting factor to atom pair recovery is now the reverse STIRAP transfer $\mathcal{F}_\mathrm{STIRAP}$ with which we scale the recovery probability of all the traps in the array.
Non-unity detection fidelities $P_d$ do not cause molecule loss but instead result in unoccupied molecule traps being inserted into the final array.
This reduces the average occupancy of ``filled'' traps while increasing the length of the array that is ``filled''.
The total number of molecules in the array is the same for all $P_d$ shown here.
These simulations suggest that a Rydberg excitation fidelity of $>84\,\%$ (well below the $\sim99\,\%$ that has been reported for Rb \cite{Evered2023}) will enable the preparation of defect-free arrays of tens of assembled  molecules.

The upgrades to our experiment described above will allow the formation of defect-free arrays of molecules comparable in size to those demonstrated with directly-cooled molecules. For comparison, stochastic loading probabilities of $\sim35\%$ are typical for an array of CaF molecules in optical tweezers~\cite{Bao2023a}.
The rearrangement of such molecules has been demonstrated to obtain defect-free arrays of up to 16 molecules with a probability $>0.6$ for a reported single-particle rearrangement fidelity of $97.4(1)\%$ and a state-preparation fidelity of $82.4(11)\%$~\cite{Holland2023}.
In this work, the probability to convert an atom pair into a ground state molecule is $48(2)\%$ and the direct detection of molecules in state \ground will allow for a rearrangement fidelity limited by the STIRAP fidelity (currently $96.4(1)\%$).
All molecules formed in this experiment occupy a single internal state.
Furthermore, for assembled molecules such as RbCs, the molecule inherits the motional state of the center of mass of the atom pair from which it is assembled.
As only atom pairs in the ground state of relative motion are converted into molecules, the formed molecules usually occupy the three-dimensional motional ground state. We estimate that this is true for $\sim66\%$ of the molecules formed in our experiment~\cite{Guttridge2023}.
This efficiency is comparable to the $54(18)\%$ occupancy of the three-dimensional motional ground state achieved after Raman sideband cooling of CaF molecules in optical tweezers \cite{Lu2023,Bao2023b}.

\section{Conclusion}
In conclusion, we have established a suite of experimental techniques for enhanced control of individual ultracold molecules assembled from ultracold atoms confined in optical tweezer arrays. We have quantified the efficiency of each step in the method used to form RbCs molecules in optical tweezers and have described an adaptable technique for detecting molecule formation errors. We have demonstrated global and local control of multiple rotational states of individually-trapped molecules and combined this with a technique for the detection of multiple rotational states in a single run of the experiment. Using mid-sequence detection of formation errors, we have demonstrated the rearrangement of assembled molecules to prepare defect-free arrays. Finally, we have discussed a feasible route to scaling to larger defect-free arrays of molecules.

The advances demonstrated here lay the foundation for new experiments in quantum science that exploit the rich internal structure and dipolar interactions of molecules~\cite{Carr2009,Bohn2017,Softley2023}. We have developed a range of techniques for the control and readout of the rotational states of molecules using optical-tweezer arrays that can be readily extended beyond the two rotational states used in this work. This extension will facilitate the realization of synthetic dimensions~\cite{Sundar2018} and qudits~\cite{Sawant2020} with molecules. Furthermore, the combination of site-resolved control of rotational states with our scheme for the detection of multiple rotational states, allows for the local shelving of molecules outside the detected rotational subspace. This capability enables mid-circuit measurements of a subset of molecules, which may be exploited to enhance precision measurements with molecules~\cite{Rosenband2013} or used for quantum information processing applications, such as measurement-based quantum computation~\cite{Raussendorf2001,Briegel2009} and quantum error correction~\cite{Shor1996,Steane1996}.

\emph{Note:} During completion of this work we became aware of related work using NaCs molecules in the Ni group at Harvard University \cite{picard2024}.

\begin{acknowledgments}
We thank Erkan Nurdun for developing the frequency stabilization system for the Rb high field imaging laser and Jonathan Pritchard for valuable discussions regarding the implementation of high field imaging. We thank Stefan Spence for performing measurements of photoassociation of the weakly bound molecular states and Olivier Dulieu for helpful discussions about likely transition candidates.
We acknowledge support from the UK Engineering and Physical Sciences Research Council (EPSRC) Grants EP/P01058X/1, EP/V047302/1, and EP/W00299X/1, UK Research and Innovation (UKRI) Frontier Research Grant EP/X023354/1, the Royal Society, and Durham University. 

The data presented in this paper are available from \cite{dataset}.
\end{acknowledgments}

\appendix

\section{Experimental methods}\label{app:experiment}
The methods used in our experiment have been extensively described in previous works \cite{Brooks21,Spence22,Ruttley2023,Guttridge2023}.
Here we provide an overview of recent upgrades relevant to this work.

\subsection{STIRAP beams}
The efficiency of one-way STIRAP transfer in our experiment has been increased from the previously reported value of 91(1)\% \cite{Guttridge2023} to the 96.4(1)\% efficiency reported here by increasing the Rabi frequencies with which we couple to state \intermediate.
This has been achieved by amplifying the pump and Stokes beams with fiber amplifiers (both supplied by Precilasers) to have powers of $\sim44$\,mW and $\sim27$\,mW focused to $1/e^2$ beam waists of $\sim80$\,$\mu$m and $\sim70$\,$\mu$m at the molecules, respectively.
With this we achieve a pump Rabi frequency of 493(5)\,kHz and a Stokes Rabi frequency of 876(4)\,kHz.
Further increase of the Rabi frequencies beyond these values reduces the efficiency of the STIRAP transfer.
We believe that the transfer efficiency at higher Rabi frequencies is limited by laser frequency noise \cite{ivanov2004,Huang2022} caused by servo bumps from frequency stabilization to a high-finesse optical cavity \cite{Leseleuc2018,Day2022}. In the future we plan to suppress this noise with feed-forward techniques \cite{li2022,chao2023}.

\subsection{High-field imaging}
As described in the main text, to detect molecule formation errors mid-sequence we must image Rb atoms initially in the state ${(5\mathrm{s}_{1/2}, f=1,m_f=1)}$ at a magnetic field of $181.699(1)$\,G. Here we give further details of the microwave adiabatic rapid passage (ARP) and subsequent resonant imaging mentioned in Sec.~\ref{sec:rearrangement}.

The ARP transfer from ${(1,1)}$ to ${(2,2)}$ is implemented using a loop antenna mounted outside the vacuum chamber approximately 12\,mm from the atoms confined in optical tweezers. With this antenna we produce microwave radiation that couples the states ${(1,1)}$ and ${(2,2)}$ with a Rabi frequency of 7.6(3)\,kHz. We use a magnetic field sweep of 72\,mG in 2.5\,ms to adiabatically transfer the atoms into ${(2,2)}$. For this Rabi frequency, the efficiency of the transfer is limited to $90\%$ due to magnetic field noise. In future experiments we plan to increase the coupling strength to improve the transfer.

We then image the Rb atoms on the closed transition ${(2,2)} \rightarrow {(3',3')}$ with resonant \sigmap polarized light with a peak intensity of 10 $I_\mathrm{sat}$. The use of the closed transition prevents atoms decaying to states that are dark to the imaging light due to the large Zeeman shifts. The imaging light is sourced from a dedicated laser which is frequency stabilized relative to the main cooling laser with a beat note lock to be resonant with the imaging transition at the operating magnetic field of $181.699(1)$\,G.
The primary limitation to the achievable imaging fidelity is the loss of Rb atoms before the number of scattered photons that are detected is sufficient to differentiate occupied traps from the background.
This loss is caused by the recoil momentum imparted by imaging photons heating atoms out of the traps.
To combat this, we increase the peak depth of the tweezers to 2.5(1)\,$\mathrm{mK}\times k_B$.
During the imaging procedure we modulate the imaging and trapping light in antiphase; this avoids light shifts caused by the deep trap which would otherwise cause broadening of the signal histograms and loss from dipole-force fluctuations~\cite{Martinez2018}.
The duty cycle of the trapping (imaging) light is approximately 80\% (10\%) and we estimate that approximately $10^{4}$ photons are scattered before the atoms are lost.  

To enhance the preparation efficiency in (2,2) beyond the $90\%$ achievable with ARP alone, and consequently improve the Rb detection fidelity, we implement additional optical pumping methods. We apply optical pumping light resonant with the ${(1,1) \rightarrow (2',2')}$ transition at 4.78\,G, but off-resonant at our magnetic field of $181.699(1)$\,G. We also apply a microwave field resonant with the ${(1,1) \rightarrow (2,2)}$ transition during imaging to continuously pump atoms from the dark state $(1,1)$ to the bright state $(2,2)$. We find these steps pump $>99\,\%$ of the atoms into the bright state for imaging.

\section{Molecular states} \label{app:mol_states}
In the presence of externally applied optical and magnetic fields we resolve the hyperfine structure within each rotational manifold in RbCs.
This hyperfine structure results from coupling between the rotational angular momentum and the nuclear spins of the constituent atoms 
($i_\mathrm{Rb} = 3/2$ and $i_\mathrm{Cs} = 7/2$).
This splits each $\ket{N,M_N}$ state into $(2i_\mathrm{Rb}+1)(2i_\mathrm{Cs}+1) = 32$ hyperfine states. 

The magnetic field in our experiment is typically $\sim$\,200\,G which is not high enough to decouple the rotational and nuclear angular momenta. 
As described in the main text, generally the only good quantum numbers that can be used to describe a given hyperfine sublevel are $N$ and $M_{F}=M_{N}+m_{\mathrm{Rb}}+m_{\mathrm{Cs}}$.
In the main text, we use the state labels $\ket{N,M_N}$ where mixed states are labeled with $M_N$ of their component state with the largest probability amplitude.
Neither of these labeling schemes is sufficient to identify a given hyperfine state uniquely.
Therefore, here we use the labeling scheme of Blackmore \emph{et al.} \cite{blackmore2020b} where states are labeled by $\ket{N,M_{F}}_{k}$. 
Here $k$ is an index enumerating states in order of increasing energy such that $k=0$ is the lowest energy state for given values of $N$ and $M_{F}$.

In Table~\ref{tab:state_compositions} we list the hyperfine states used in this work.
We give the label $\ket{N,M_N}$ used in the main text, the label $\ket{N,M_F}_k$ following the scheme of Blackmore \emph{et al.}, and the full state composition in the $\ket{N,M_N,m_\mathrm{Rb},m_\mathrm{Cs}}$ basis.
The state compositions are calculated for a molecule in a 1065.512\,nm tweezer of intensity $I_{1066}=3.07$\,kW/cm$^2$ at a magnetic field of $181.699$\,G using the molecular constants and polarizabilities as described in the main text.
The components of the $N=1$ states to which we can couple with microwave radiation from state \ground have $m_\mathrm{Rb} = 3/2$ and $m_\mathrm{Cs} = 7/2$ and are highlighted in bold.

\begin{table}[b]
    \centering
        \caption{Rotational and hyperfine states of RbCs used in this work. We give the $\ket{N,M_N}$ label used in the main text, the corresponding $\ket{N,M_F}_k$ label, and the state compositions in a 1065.512\,nm tweezer of intensity $I_{1066}=3.07$\,kW/cm$^2$ at a magnetic field of 181.699\,G. The state components with $m_\mathrm{Rb} = 3/2$ and $m_\mathrm{Cs} = 7/2$ are shown in bold.}
        \begin{tabular}{l|l|l}
        $\ket{N,M_N}$ & $\ket{N,M_F}_k$ & $\ket{N,M_N,m_\mathrm{Rb},m_\mathrm{Cs}}$ \\
        \hline
        \hline
        $\ket{0,0}\equiv$\ground & $\ket{0,5}_0$ & $\mathbf{\ket{0, 0, 3/2, 7/2}}$\\
        \hline
        $\ket{1,-1}$ & $\ket{1,4}_1$ &  $\mathbf{-0.827\ket{1, -1, 3/2, 7/2}}$\\
        & & \xspace$-0.438\ket{1, 0, 1/2, 7/2}$\\
        & & \xspace$+ 0.293\ket{1, 1, -1/2, 7/2}$\\
        & & \xspace$-0.196\ket{1, 0, 3/2, 5/2}$\\
        & & \xspace$-0.019\ket{1, 1, 3/2, 3/2}$\\
        & & \xspace$+ 0.011\ket{1, 1, 1/2, 5/2}$\\
        \hline
        $\ket{1,0}$ & $\ket{1,5}_0$ & $\mathbf{-0.971\ket{1, 0, 3/2, 7/2}}$\\
        & & \xspace$+ 0.236\ket{1, 1, 1/2, 7/2}$\\
        & & \xspace$-0.039 \ket{1, 1, 3/2, 5/2}$\\
        \hline
        $\ket{1,1}$ & $\ket{1,6}_0$ & $\mathbf{\ket{1, 1, 3/2, 7/2}}$\\
        \hline
        $\ket{2,2}$ & $\ket{2,7}_0$ & $\mathbf{\ket{2, 2, 3/2, 7/2}}$\\
        \hline
        $\ket{3,3}$ & $\ket{3,8}_0$ & $\mathbf{\ket{3, 3, 3/2, 7/2}}$
    \end{tabular}

    \label{tab:state_compositions}
\end{table}


%

\end{document}